\documentclass[aps,showpacs,twocolumn,superscriptaddress]{revtex4}

\usepackage{dcolumn}   %for decimal-point aligned tables
\usepackage{graphicx}
\usepackage{amsfonts}
\usepackage{amsmath,amssymb}
\usepackage{bm}
\usepackage{color} %used for font color

\newcommand{\beq}{\begin{equation}}
\newcommand{\eeq}{\end{equation}}
\newcommand{\bea}{\begin{eqnarray}}
\newcommand{\eea}{\end{eqnarray}}
\newcommand{\ben}{\begin{eqnarray*}}
\newcommand{\een}{\end{eqnarray*}}

\newcommand{\kvec}{{\bf k}}

\newcommand{\abinitio}{\textit{ab initio}}

\newcommand{\Rvec}{{\bf R}}
\newcommand{\kf}{k_{\text{F}}}

\begin{document}

\title{Testing the density matrix expansion against
   ab initio \\ calculations of trapped neutron drops}
   %: Model calculations}

\author{S.K.~Bogner}
\affiliation{National Superconducting Cyclotron Laboratory and Department of 
Physics and Astronomy, Michigan State University, East Lansing, MI 48824}

\author{R.J.~Furnstahl}
\affiliation{Department of Physics, Ohio State University, Columbus, OH 43210}

\author{H. Hergert}
\affiliation{National Superconducting Cyclotron Laboratory and Department of 
Physics and Astronomy, Michigan State University, East Lansing, MI 48824}

\author{M.~Kortelainen}
\affiliation{Department of Physics \& Astronomy, University of Tennessee, Knoxville, Tennessee 37996 \\
Physics Division,  Oak Ridge National Laboratory, Oak Ridge, Tennessee 37831}

\author{P.~Maris}
\affiliation{Department of Physics and Astronomy, Iowa State University, 
              Ames, IA\ 50011}

\author{M.~Stoitsov}
\affiliation{Department of Physics \& Astronomy, University of Tennessee, Knoxville, Tennessee 37996 \\
Physics Division,  Oak Ridge National Laboratory, Oak Ridge, Tennessee 37831}
%\affiliation{INRNE, Bulgarian Academy of Sciences, Sofia-1784, Bulgaria}

\author{J.P.~Vary}
\affiliation{Department of Physics and Astronomy, Iowa State University, 
              Ames, IA\ 50011}

\date{\today}

\begin{abstract}
Microscopic input to a universal nuclear energy density functional
can be provided through the density matrix expansion (DME), which has
recently been revived and improved.
Several DME  implementation strategies are tested for neutron drop
systems in harmonic traps 
by comparing to Hartree-Fock (HF) and \abinitio\ no-core full
configuration (NCFC) calculations with
a model interaction (Minnesota potential).
The new DME with exact treatment of Hartree contributions
is found to best reproduce HF results and supplementing
the functional with fit Skyrme-like contact terms shows
systematic improvement toward the full NCFC results.
\end{abstract}


\maketitle

\section{Introduction}
\label{sec0}

Experiments at
radioactive ion beam facilities, studies of astrophysical systems such as neutron stars and supernovae, and nuclear energy and security needs have motivated multipronged efforts to develop nuclear energy 
density functionals (EDF's) with substantially reduced errors and improved predictive power away from stability. While great progress has been made in extending the reach of \abinitio\ wave function methods beyond the lightest nuclei~\cite{Pieper:2004qh,Quaglioni:2007qe,Hagen:2008iw,Maris:2009bx}, the EDF approach remains the most computationally feasible method for a comprehensive description of medium and heavy nuclei~\cite{Bender:2003jk}.
However, the \abinitio\ methods are vital tools for reaching the
goal of robust functionals informed by microscopic internucleon interactions.
As part of an ongoing program to achieve this goal, 
 in this paper we investigate trapped neutron drops with a model interaction.
In particular, EDF calculations using several
density matrix expansion (DME) implementations are confronted with
ab initio no-core full configuration (NCFC)~\cite{Maris:2008ax}
results.  

The comparisons presented here exploit developments achieved within
the Universal Nuclear Energy Density Functional (UNEDF) SciDAC-2 
collaboration~\cite{unedf:2007,unedfweb}.
The UNEDF project aims to develop a comprehensive theory
of nuclear structure and reactions utilizing the most advanced computational resources and algorithms available, including high-performance computing techniques to scale to petaflop platforms and beyond~\cite{unedf:2007}.
One element of the UNEDF program involves the direct injection of microscopic
physics into novel energy functionals, 
with the DME a key tool~\cite{Bogner:2008kj,Drut:2009ce,Gebremariam:2009ff,Gebremariam:2010ni}.
Another element has led to 
efficient density functional theory (DFT) solvers adapted for the DME~\cite{Stoitsov:2010ha}
and to neutrons in external traps, which allow accurate and rapid testing
of candidate functionals~\cite{Kortelainen:2010hv}.
A third element is the extensive development of \abinitio\ methods,
including improved computational efficiencies~\cite{Maris:2010aaa} 
and extrapolation techniques for the NCFC~\cite{Maris:2008ax} that allow exact calculations
(with error bars) of the same neutron
drop systems to which the functionals are applied.
 
Neutron drops 
are a powerful theoretical laboratory for improving
existing nuclear energy functionals.
Microscopic input to EDF's is particularly needed for neutron-rich
nuclei, where there are fewer constraints from experiment and where
novel density dependencies 
from long-range pion exchange
% in two- and three-body forces 
may be particularly important to incorporate.
The properties of uniform neutron matter have been used in the past
as a constraint on phenomenological functionals (e.g., see Refs.~\cite{Chabanat:1997qh,AlexBrown:1998zz}), but computational advances now
allow accurate microscopic many-body calculations
of inhomogeneous neutron drops in external potentials using quantum Monte Carlo or NCFC methods~\cite{Gandolfi:2010za,Carlson:2011aa}.
These calculations can be used to identify deficiencies in existing functionals
(e.g., of the Skyrme type as in Ref.~\cite{Gandolfi:2010za}), to suggest or calibrate new versions, or simply to provide control data that supplements experiment for the optimization of functionals.

Neutron drops also provide favorable environments for the development 
and testing of
non-empirical (i.e., microscopically based) functionals.
The necessity of an external potential (because the untrapped system is unbound,
with positive pressure) is turned into a virtue
by allowing external control over the environment.
Density Functional Theory (DFT), 
which provides the theoretical underpinning and computational framework for building a nuclear EDF, dictates that the same functional applies
for any external potential, which can therefore be varied to probe
and isolate different aspects of the EDF.
In contrast, the treatment of self-bound systems (such as ordinary nuclei)
has much less flexibility.
Furthermore, there are serious
complications from symmetry breaking~\cite{Duguet:2010cv},
particularly for the 
relatively small systems where \abinitio\ methods can also be applied.

The density matrix expansion was introduced long ago by Negele and
Vautherin, who applied it to G-matrix effective interactions to derive
a Skyrme-like Hartree-Fock (HF) energy density functional.
The DME has been revisited in recent years with the goal of applications 
to nuclear interactions sufficiently soft that many-body perturbation theory (MBPT) for nuclei is a quantitative framework.
There are new formal DME developments,
as well as new formulations that include hybrids between purely
\abinitio\ and phenomenological functionals.
These must be tested and validated (or discarded if found to be inadequate); in general, due to the complexity of the various DME procedures and the involved density dependence they generate for the subsequent functional, we have no \textit{a priori} guidance for the accuracy. 

In this paper, we isolate the DME issues by using a simplified model
interaction, the two-body Minnesota potential.  Focusing on neutron
drops is also advantageous because the self-consistent solution of a self-bound system magnifies approximation errors to the extent that it is difficult
to analyze them.
Of course, the main drawback of using neutron drops is the absence of control by experimental data.  In our case this is not so relevant as we use a simple model interaction sufficient for conducting our basic tests of these methods.
Our results here provide a foundation for testing more realistic
interactions and improved functionals (e.g., that include pairing), and
for extensions to self-bound nuclei.

In Section~\ref{sect:background},
we briefly review the methods and inputs used.  Results for the Minnesota potential are given in Section~\ref{sect:results}  for representative
trap potentials and two closed-shell neutron drop systems.
We summarize our observations and discuss the next steps to take in Section~\ref{sect:summary}.

\section{Background}
\label{sect:background}

The present calculations combine ingredients from several
parts of the UNEDF project.  The technical details are described in full
elsewhere, so we merely review the essential features.

\subsection{DME}
 \label{subsect:DME}

The density matrix expansion
(DME) introduced by Negele and
Vautherin~\cite{Negele:1972zp,Negele:1975zz}
provides a route to an EDF
based on microscopic nuclear interactions through a quasi-local
expansion of the energy in terms of various densities:
$\rho(\Rvec)$, $\tau(\Rvec)$, $\nabla^2\rho(\Rvec)$, 
and so on.
Kohn-Sham single-particle potentials are immediately obtained from simple
functional derivatives with respect to these densities.
The DME originated as an expansion of the Brueckner-Hartree-Fock energy
constructed using the nucleon-nucleon 
G-matrix ~\cite{Negele:1972zp,Negele:1975zz}, which was
treated in a local (i.e., diagonal in coordinate representation) approximation.

The DME has been reformulated for spin-saturated nuclei
using non-local low-momentum interactions in 
momentum representation~\cite{Bogner:2008kj}, 
for which G-matrix summations are
not needed because of the softening of the interaction.
When applied to a Hartree-Fock energy functional, the DME yields
an EDF in
the form of a generalized Skyrme functional that is compatible
with existing codes, by replacing 
Skyrme coefficients with density-dependent functions.
As in the original application, a key feature of the DME is
that it is not a pure short-distance expansion but includes
resummations that treat long-range interactions correctly
in a uniform system.

Extensions of the first calculations from Ref.~\cite{Bogner:2008kj}
have modified
the DME formalism from Negele and Vautherin~\cite{Negele:1972zp},
which provides an extremely poor description of the 
vector part of the density matrix~\cite{Gebremariam:2009ff}. 
The standard DME is much better at reproducing the scalar
density matrices, but errors are still sufficiently large that the
discrepancy with full finite-range Hartree-Fock calculations can 
reach the MeV per particle level. 
Gebremariam and collaborators~\cite{Gebremariam:2009ff} introduced a new phase space averaging (PSA) approach that accounts for the diffuse Fermi
surface~\cite{Durand:1982aa} and anisotropy~\cite{Bulgac:1996aa} of the local momentum distribution, with no free parameters.
The PSA treatment leads to substantial improvements, particularly for the
vector density matrices, where
relative errors in integrated quantities are reduced by as much as an order of
magnitude across isotope chains~\cite{Gebremariam:2009ff}. 
In the present work, we test the difference between the original
Negele-Vautherin (NV) and the new PSA prescriptions only for scalar parts.

The Fock energy exhibits spatial non-localities due to the convolution of finite-range interaction vertices with non-local density matrices. The DME factorizes the non-locality of the one-body density matrix 
by expanding it into a finite sum of terms that are separable in relative $\bm{r}\equiv \bm{r}_1-\bm{r}_2$ and center of mass $\bm{R}\equiv (\bm{r}_1+\bm{r}_2)/2$ coordinates. 
For example, in notation introduced in
Refs.~\cite{Gebremariam:2009ff,Gebremariam:2010ni}, one expands the spin-scalar part (in both isospin channels) of the one-body density matrix as
\beq
\rho_{t} (\bm{r}_1, \bm{r}_2)  \approx  \sum^{n_{\text{max}}}_{n=0} \Pi_n (k  r)
   \,\, {\cal P}_n (\bm{R}) \;, \label{approxscalar}
%\bm{s}_t (\bm{r}_1, \bm{r}_2) &\approx& \sum^{m_{\text{max}}}_{m=0} \Pi_m (k r)
%   \,\, {\cal Q}_m (\bm{R}) \;,\label{approxvector}
\eeq
where the functions $\{{\cal P}_n (\bm{R})\}$ denote various local densities 
and their gradients (through second order in the present work) and
$\Pi_n (k r)$ denotes the so-called
$\Pi-$functions, which depend on the particular formulation of the DME
(here NV or PSA). 
The arbitrary momentum $k$ sets the scale for the decay in the off-diagonal direction.
As in Ref.~\cite{Stoitsov:2010ha}, we will take  $k$ to be the local Fermi momentum related to the isoscalar density through
\begin{equation}
   k \equiv k_F(\bm{R}) = \biggl(\frac{3\pi^2}{2}\rho_0(\bm{R})\biggr)^{1/3} \;,
   \label{LDA}
\end{equation}
although other choices are possible that include additional $\tau$ and $\Delta\rho$ dependencies~\cite{Campi:1978}.  

It is possible to apply the DME to both Hartree and Fock energies so that the
complete Hartree-Fock energy is mapped into a local functional.  From the
earliest DME work, however, it was found that treating the Hartree contributions
exactly provides a better reproduction of the density fluctuations and the
energy produced from an exact HF calculation \cite{Negele:1975zz,Sprung:1975}.
Restricting the DME to the exchange contribution significantly reduces the
self-consistent propagation of errors~\cite{Negele:1975zz}. In addition,
treating the Hartree contribution exactly does not complicate the numerical
solutions of the resulting self-consistent equations compared to applying the
DME to both Hartree and Fock terms.  Here we will compare several prescriptions
for handling the Hartree contribution,
including a Taylor series expansion~\cite{Carlsson:2010da}.

A consistent and systematic extension of the DME procedure beyond the Hartree-Fock level of MBPT has yet to be formulated. 
For now, attempts to microscopically construct a {\it quantitative} Skyrme-like EDF  use some {\it ad hoc} approximations (e.g., using averaged rather than state-dependent energy denominators) when applying the DME to iterated contributions beyond the HF level and/or re-introduce some phenomenological parameters to be adjusted to data~\cite{Negele:1972zp,Negele:1975zz,Hofmann:1997zu,Kaiser:2002jz,Kaiser:2009me}.
Recent work based on chiral NN and NNN
effective field theory interactions motivates an approach to building upon
the DME/HF functional that incorporates important microscopic physics
while exploiting the highly developed Skyrme EDF technology~\cite{Gebremariam:2009ff,Gebremariam:2010ni,Stoitsov:2010ha}.

The structure of the chiral interactions is such that each coupling in 
the DME/HF
functional is decomposed into a density-independent coupling constant arising from
zero-range contact interactions and a coupling function of the
density  arising from the universal long-range pion exchanges. 
This clean separation between long- and short-distance physics at the HF level
suggests a semi-phenomenological approach where the long-distance couplings ($g_t^{m}(\Rvec;V_{\pi})$) are kept as is, and the zero-range constants
$C^{m}_t$ are optimized to finite nuclei and infinite nuclear matter properties~\cite{Gebremariam:2009ff,Gebremariam:2010ni}. 
Thus,
\beq
g^{\rho\tau}_t  \equiv  g^{\rho\tau}_t(\Rvec;V_{\pi}) + C^{\rho\tau}_t(V_{{\rm ct}}) \;,
\eeq
and so on, so that the DME functional splits into two terms,
\begin{equation}
 E[\rho]=E_{\rm ct}[\rho]+E_{\pi}[\rho] \;,
  \label{dme2}
\end{equation}
where the first term $E_{\rm ct}[\rho]$ collects all contributions from the contact part of the interaction plus higher-order short-range contributions encoded through the optimization to nuclei and nuclear matter, while the second term $E_{\pi}[\rho]$ collects  the long-range NN and NNN pion exchange contributions at the Hartree-Fock level.

Because the contact contributions have essentially the same structure as those
entering empirical Skyrme functionals, a microscopically guided Skyrme
phenomenology has been suggested in which the contact terms in the DME
functional are released for optimization to finite-density observables~\cite{Gebremariam:2009ff,Gebremariam:2010ni}. 
This empirical procedure is supported by the observation that the dominant bulk correlations in nuclei and nuclear matter are primarily short-ranged in nature, as evidenced by Brueckner-Hartree-Fock (BHF) calculations where the Brueckner G-matrix ``heals'' to the free-space interaction at sufficiently large distances.
One can loosely interpret the refit of the Skyrme constants to data as approximating the short-distance part of the G-matrix with a zero-range expansion through second order in gradients.
In doing
so, the constants can capture short-range correlation energy contributions beyond
Hartree-Fock. We will test this strategy for incorporating
BHF correlations from the Minnesota potential, with
the free parameters
of the volume part of the functional fixed to properties of infinite
neutron matter and the free surface parameter adjusted to NCFC results. 
We will also consider a direct density-dependent modification to
model BHF correlations.

\subsection{Minnesota Potential}

All of the calculations reported here use the Minnesota potential.
This is a local NN-only potential that is the sum of
three Gaussians in the radial coordinate $r_{ij}$~\cite{Thompson:1977zz}:
\beq
  V_{ij} = [V_R + \frac12(1+P^\sigma_{ij})V_t
    + \frac12(1-P^\sigma_{ij})V_s]
    \frac12(1 + P^r_{ij}) 
    \;,
    \label{eq:Vij}
\eeq
where $P^\sigma$ and $P^r$ are spin and space exchange operators, respectively,
and 
\bea
  V_R &=& V_{0R} e^{-\kappa_R r^2_{ij}} \;,
  \label{eq:VR}
  \\
  V_t &=& -V_{0t} e^{-\kappa_t r^2_{ij}} \;,
  \label{eq:Vt}
  \\
  V_s &=& -V_{0s} e^{-\kappa_s r^2_{ij}} \;.
  \label{eq:Vs}
\eea
The parameters defining the $V_{ij}$ are given in Table~\ref{tab:minpot}.
(Note: we have taken the exchange-mixture parameter $u$ in Ref.~\cite{Thompson:1977zz} equal to be one.)
The Minnesota potential reproduces NN effective range parameters
and gives reasonable results for the binding energies of light nuclei.
It is often used as a semi-realistic potential in model calculations.

\begin{table}[h]
\caption{Parameters defining the Minnesota potential, see Eqs.~\eqref{eq:Vij}--\eqref{eq:Vs}.}
\label{tab:minpot}
\begin{tabular}{cc|cc}
  \hline
  $V_\alpha$  & value  & $\kappa_\alpha$ & value \\
  \hline
  $V_{0R}$ & 200.0\,MeV & $\kappa_R$ & $1.487\,\text{fm}^{-2}$ \\
  $V_{0t}$ & 178.0\,MeV & $\kappa_t$ & $0.639\,\text{fm}^{-2}$ \\
  $V_{0s}$ & \phantom{1}91.85\,MeV & $\kappa_s$ & $0.465\,\text{fm}^{-2}$ \\
  \hline
\end{tabular}
\end{table}

For our purposes, the important characteristics of this potential 
are that it is local,
which makes possible the immediate adaptation of current
DME technology (which is not fully developed for non-local potentials),
and that it is moderately soft, so that HF is a reasonable starting
point and convergence is adequate in the NCFC.
Because we plan to use low-momentum interactions in the future, this
softness is consistent rather than a shortcoming of the model.
While the Minnesota potential lacks important features of realistic
interactions, such as tensor forces and three-nucleon interactions,
it provides a convenient, non-trivial test case for the DME that sets the
stage for future tests.

\subsection{EDF Solvers}

The DME-based functionals described in the last section and in
Ref.~\cite{Stoitsov:2010ha} have been implemented in the DFT
solvers HFBRAD~\cite{Bennaceur:2005mx} and HFBTHO~\cite{Stoitsov:2004pe}.
HFBRAD is a very fast solver for spherical nuclei and density-dependent
local density approximations, while HFBTHO is much slower but calculates
spherical and axially deformed nuclei, and can handle additional
gradient corrections.  
A Fortran module for both solvers has been developed 
to implement the density-dependent parts of the EDF from the DME applied to chiral effective potentials~\cite{dmemodule}. 
The module contains all of the lengthy  expressions for the DME couplings and 
their functional derivatives with respect to the density matrix, 
and for numerically stable approximations. 
The module also has the capability to calculate related 
infinite nuclear matter properties.
We have developed a similar module that
can handle expressions coming from the DME of the Minnesota potential plus  external potentials.
This module was linked
to existing DFT solvers to calculate results presented here.

\subsection{NCFC}

To test the DME calculations, we use the \textit{ab initio}
no-core full configuration (NCFC) 
approach~\cite{Maris:2008ax,Bogner:2007rx} to provide exact
results (with errors bars).
The NCFC is closely related to the no-core shell model 
(NCSM)~\cite{Zheng:1994jr,Zheng:1995td,Navratil:2000ww,Navratil:2000gs},
as both employ a many-body harmonic oscillator basis that treats all
nucleons as spectroscopically active.
The basis space includes all many-body states with excitation
quanta less than or equal to $N_{\rm max}$.
A general feature is the possibility to completely remove spurious
center-of-mass excitations, but the present application with an
external potential does not exploit this capability.  Rather,
the center-of-mass motion is part of the physical system.
 The main difference between the NCFC and NCSM is that the NCSM employs an interaction renormalized to the finite many-body basis, such as the Lee-Suzuki effective interaction.

The NCFC approach involves the extrapolation of a sequence of finite matrix 
results with the bare interaction 
(as opposed to a Lee-Suzuki effective interaction) 
to the infinite basis space limit.  This makes it possible to obtain basis-space-independent results for binding energies 
and other observables, and to evaluate their numerical uncertainties.
The extrapolation methods are described in Ref.~\cite{Maris:2008ax}.
A recent calculation of $^{14}$F in Ref.~\cite{Maris:2009bx},
made prior to the first experimental measurements,
illustrates the predictive power of the NCFC approach when coupled
with leadership class computer resources.
Note that the present calculations do not exploit the full capabilities of the
codes and computers available to further minimize theoretical errors, 
because current error bars are small enough for the present application. 

To solve for ground-state energies, radii, and form factors of trapped
neutron systems,
the code MFDn~\cite{Vary:1992mfd,Sternberg:2008aa,Vary:2009qp,Maris:2010aa} was generalized to allow for external potentials. 
Only spherically symmetric harmonic oscillator traps are used in the
current investigation, but other shapes and deformed traps
are also directly available.

\section{Results}
\label{sect:results}

The neutrons are confined by an external single-particle harmonic
potential:
\beq
  v_{\rm ext}(r) = \frac12 m\Omega^2 r^2 \;,
\eeq
with harmonic oscillator parameter $\hbar\Omega$ varied from
5\,MeV to 20\,MeV.
The calculations here use
$N=8$ and $N=20$ neutrons, which form closed shells.  In the future
we will revisit this problem with pairing included and consider
intermediate $N$ values.
Accurate NCFC results are limited to larger oscillator parameters because
of slow convergence with $N_{\rm max}$ for the Minnesota potential.
(Note: quantum Monte Carlo techniques such as GFMC or AFDMC
are effective for smaller $\hbar\Omega$ and could be used for additional
comparisons.)
In some cases, extrapolations are not reliable and so only upper
bounds to the total energy are given.

Comparisons between different DME treatments of the Hartree term
are given in Tables~\ref{tab:HartreeE}, \ref{tab:HartreeInt} and \ref{tab:Hartreer}.
The full HF results provide a baseline for comparison of
the Negele-Vautherin (HF/NV) and Phase-Space-Averaging (HF/PSA) approximations
to HF,
with variations based on how the Hartree part of the DME is treated.
For each of the two DME implementations, there
are three possibilities:  treat Hartree with the same DME (NV or PSA), use
a naive Taylor expansion (NT), or treat it exactly.
We split the total energy into internal and trap contributions, with
\beq
  E_{\rm int} = E_{\rm tot} - U_{\rm ext}
\eeq
and
\beq
  U_{\rm ext} = 4\pi \int\! dr\, r^2 v_{\rm ext}(r)\,\rho(r)
  \;.
\eeq
Results are presented as deviations  from the full HF results
of the total and internal energies and the radii.
These results are for spherical solutions, which
were shown to minimize the energy. 
That is, by imposing a non-zero quadrupole moment as a constraint, we found 
in all cases that the total energy rapidly increases as the
quadrupole moment deviates from zero.

\begin{table}[tb]
\caption{Comparison of DME approximations to HF
total energies for different treatments
of the Hartree term, expressed as deviations from the full
Hartree-Fock results  $\Delta E_i = E_i - E_{\rm HF}$ in MeV.}
 \label{tab:HartreeE}
 \smallskip
 \begin{tabular}{cc|rrr|rrr}
   \hline
      &               &  \multicolumn{3}{c|}{HF/NV} &
                         \multicolumn{3}{c}{HF/PSA} \\
  $N$ & $\hbar\Omega$ &  NV  & NT  & exact  & PSA & NT  & exact \\ 
  \hline
   8  & 3   & 0.1   & 0.2    & 0.1   & $0.0$  & 0.1  & 0.0  \\
   8  & 5   & 0.4   & 0.8    & 0.4   & $-0.1$  & 0.6  & 0.2  \\
   8  & 10  & 2.1   & 5.1    & 2.0   & $-1.7$  & 4.1  & 0.9   \\
   8  & 15  & 4.2   & 12.9   & 4.6   & $-7.1$  & 10.8 & 2.1  \\
   8  & 20  & 6.0   & 24.2   & 7.7   &         & 20.9 & 3.4   \\
   \hline
  20  & 3   & 0.5   & 0.8   & 0.6  & $-0.1$    & 0.4 & 0.2   \\
  20  & 5   & 1.8   & 3.4   & 2.3  & $-1.0$    & 2.0 & 0.9   \\
  20  & 10  & 5.9   & 18.5   & 11.0  & $-14.0$ & 12.0 & 3.9   \\
  20  & 15  & 3.8   & 44.3   & 22.7  &         & 31.6 & 7.9   \\
  20  & 20  & $-17.8$ & 80.0 & 34.8  &         & 61.3 & 12.5   \\
  \hline
 \end{tabular}
\end{table}

\begin{table}[tb]
\caption{Comparison of DME approximations to HF
internal energies for different treatments
of the Hartree term, expressed as deviations from the full
Hartree-Fock results  $\Delta E_i = E_i - E_{\rm HF}$ in MeV.}
 \label{tab:HartreeInt}
 \smallskip
 \begin{tabular}{cc|rrr|rrr}
   \hline
      &               &  \multicolumn{3}{c|}{HF/NV} &
                         \multicolumn{3}{c}{HF/PSA} \\
  $N$ & $\hbar\Omega$ &  NV  & NT  & exact  & PSA & NT  & exact \\ 
  \hline
   8  & 3  & $0.0$ & $-0.0$ & $0.0$  & 0.1  & $0.0$  & $0.1$  \\
   8  & 5  & $-0.1$ & $-0.4$ & $-0.1$  & 0.2  & $-0.3$  & $-0.1$  \\
   8  & 10  & 0.1   & $-1.0$ & $-0.2$  & 1.3  & $-1.0$  & $-0.1$   \\
   8  & 15  & 1.1   & $-1.5$   & 0.3   & 5.4  & $-1.8$  &  0.1   \\
   8  & 20  & 3.2   & $-1.9$   & 1.1   &      & $-2.7$  &  0.5   \\
   \hline
  20  &  3  & $-0.2$ & $-0.4$  & $-0.3$  & 0.1 & $-0.3$ & $-0.1$   \\
  20  &  5  & $-0.3$ & $-1.2$  & $-0.6$  & 1.0 & $-0.8$ & $-0.2$   \\
  20  & 10  & 3.3   & $-2.2$   & 0.3  & 11.9 & $-2.6$ & 0.2   \\
  20  & 15  & 16.9   & $-2.0$   & 4.4  &      & $-5.4$ & 1.4   \\
  20  & 20  & 68.3  & $-1.2$   & 11.1  &      & $-8.7$ & 3.2   \\
  \hline
 \end{tabular}
\end{table}

\begin{table}[h]
\caption{Comparison of DME approximations to HF rms radii for different treatments
of the Hartree term, expressed as deviations from the full
Hartree-Fock results $\Delta r_i = r_i - r_{\rm HF}$ in fm.}
 \label{tab:Hartreer}
 \smallskip
 \begin{tabular}{cc|rrr|rrr}
   \hline
      &               &  \multicolumn{3}{c|}{HF/NV} &
                         \multicolumn{3}{c}{HF/PSA} \\
  $N$ & $\hbar\Omega$ &  NV\ \null  & NT\ \null  & exact  & PSA & NT  & exact \\
  \hline
   8  & 3   & 0.01  & \ 0.02 & 0.00  & $-0.02$   & 0.01 & $-0.01$   \\
   8  & 5   & 0.03  & \ 0.07 & 0.03  & $-0.01$   & 0.05 & 0.01   \\
   8  & 10  & 0.04   &  0.11 &  0.04 & $-0.06$   & 0.09 & 0.02   \\
   8  & 15  & 0.03   & 0.14  & 0.04  & $-0.13$   & 0.12 & 0.02  \\
   8  & 20  & 0.02   & 0.16  & 0.05  &           & 0.15 & 0.02  \\
   \hline
  20  &  3  & 0.02   & 0.05   & 0.03 & $-0.02$   & 0.02 & 0.01   \\
  20  &  5  & 0.04   & 0.09   & 0.06 & $-0.04$   & 0.06 & 0.02   \\
  20  & 10  & 0.02   & 0.13   & 0.07 & $-0.18$   & 0.09 & 0.02   \\
  20  & 15  & $-0.04$   & 0.16   & 0.07   &      & 0.13 & 0.03  \\
  20  & 20  & $-0.20$   & 0.18   &  0.05  &      & 0.15 & 0.02   \\
  \hline
 \end{tabular}
\end{table}

It is evident that the DME with PSA and exact Hartree is
systematically the closest to HF energies and radii.
For all DME approximations, internal energies are generally closer
to full HF than the total energies.  This can be understood
because the internal energy comes mostly from the volume part of the EDF.
DME and full HF agree for infinite (uniform) matter, so similar results
can be expected from this part.  
However, one 
can see that the DME approximations give slightly larger radii, which 
implies slightly larger external energies, and so larger
total energies. 
A possible explanation is that the DME coupling for 
the $\rho\Delta\rho$ term may not take into account all the surface effects.

Some of the DME/PSA entries in Tables~\ref{tab:HartreeE} and \ref{tab:Hartreer} 
are missing. In these cases, the calculation failed
to converge because of EDF instabilities~\cite{Kortelainen:2010wb}. The instabilities
occurred at high values of $\hbar\Omega$ when both the Hartree and Fock terms were taken from the DME. This could indicate some problems 
at high density in the Hartree part of the DME expressions.
These instabilities are not, however, just simply related 
to the infinite neutron matter properties, and
are therefore more involved~\cite{Kortelainen:2010wb}.

An alternative to the DME for microscopically based EDF's 
uses the more completely microscopic 
but computationally far more intensive
Optimized Effective Potential (OEP) method~\cite{Drut:2009ce}.
In Ref.~\cite{Drut:2011te}, the OEP was applied to the same model
problem of the Minnesota potential for neutrons in a trap.  Comparisons
made to exact HF results show that the exact exchange version of
OEP is almost indistinguishable from HF, in contrast to
the small but non-negligible discrepancies found here.
Future comparisons as both methods continue to be refined will help to gauge the accuracy of DME approximations and guide the development of corrections.

To test schemes for incorporating correlations beyond HF,
we use BHF calculations of neutron matter,
which we expect to be quite accurate for the Minnesota potential.  Two strategies are considered,
following the discussion in Section~\ref{subsect:DME}.

The first strategy is based on the empirical observation that the
ratio of the neutron matter HF and BHF results is a rather smooth
function of density, which we denote $f(\kf)$.  By assuming
a rank-one separable expansion of the potential, $V(\kvec,\kvec{}')$, the G-matrix would
take the form $G(\kvec,\kvec{}') \sim V(\kvec,\kvec{}')/f(\kf)$ and then taking a simple
Gaussian for the potential form factor and expanding out the
integral that appears in the definition of $f(\kf)$, one motivates
the form
  \beq
     f = a + b \rho^{1/3} + c\rho + d\rho^{5/3} + \ldots \;.
  \eeq
The coefficient $a$, $b$, $c$, and $d$ are determined by a fit.
(Calculations omitting $d$ were also made and yield similar results except
for the densest neutron drops.)  
This strategy is implemented in the DFT solvers by evaluating  
the $\rho$ dependence in the Fock terms as $\rho \rightarrow \rho(R)$, which means the DME/HF couplings simply get scaled by $f(\rho)$.
In the exact Hartree treatment, the prescription
$\rho \rightarrow 1/2 [\rho(r_1)+\rho(r_2)]$ is used and otherwise
$\rho(R)$ is used.
This approach is labeled BHF/PSA (or just BHF) 
in the subsequent tables and figures
(only results based on exact-Hartree, DME/PSA are given).

The second strategy follows Ref.~\cite{Stoitsov:2010ha} to incorporate BHF correlations by adding a contact
part to the HF functional for the Minnesota potential.
In general,
the contact part $E_{\rm ct}[\rho]$ of the EDF has the form of the standard Skyrme functional
\begin{equation}
{\cal H}_{\rm ct}(\bm{r}) = \frac{\hbar^2}{2m}\tau_0+{\cal H}^{\rm ct}_{0}(\bm{r})
  +{\cal H}^{\rm ct}_{1}(\bm{r}) \;,
  \label{cp}
\end{equation}
where
\begin{eqnarray}
  {\cal H}^{\rm ct}_{t}(\bm{r}) &=& \bigl(C_{t0}^{\rho^2}
   +C_{tD}^{\rho^2} \rho_0^{\gamma}\bigr)\rho_t^2 + C_t^{\rho\tau} \rho_t\tau_t
  + C_t^{\rho\Delta\rho} \rho_t\Delta\rho_t
  \nonumber \\
  & & \null +C_t^{\rho\nabla J} \rho_t \nabla J_t
    + C_t^{J^2} J_t^2 \;,
\label{cpt}
\end{eqnarray}
and the isospin index $t=\{0,1\}$ labels isoscalar and isovector densities, respectively.
In analogy to Ref.~\cite{Stoitsov:2010ha}, the neutron coupling constants 
$C^{i}=C^{i}_{0}+C^{i}_{1},\, i=\{\rho^{2},\rho\tau\}$ 
are fitted to reproduce the neutron matter BHF results. This allows us to constrain 
the zero-range volume parameters of the DME-based functional, but not the parameters 
entering the surface part of the functional.

We optimize the surface parameters in the DME functional in a
manner similar to the  optimization done for standard Skyrme functionals. One
could think of procedures based on semi-infinite neutron properties, or on the
leptodermous expansion of the functional~\cite{Reinhard:2005nj}. 
Here we optimize the surface coupling constant
$C^{\rho\Delta\rho}=C^{\rho\Delta\rho}_{0}+C^{\rho\Delta\rho}_{1}$ 
to NCFC $E_{\rm tot}$ values presented in Tables \ref{tab:eightN} and \ref{tab:twentyN}
by using theoretical error bars as weights. A simple minimization of the root-mean-square
deviation yields almost the same result.
The values for the neutron parameters of the Minnesota model are:
\begin{align}
  & C^{\rho^2} = -18.25 \,{\rm MeV\,fm^{3}}\;,
  \quad
  C^{\rho\tau} = 4.57 \,{\rm MeV\,fm^{5}}\;, \nonumber \\
  & C^{\rho\Delta\rho}= -1.8 \,{\rm MeV\,fm^{5}}\;, \label{eq:fitCset}
\end{align}
with $C^{\rho^2}_D=C^{\rho\nabla J} = C^{J^2} = 0$. Calculations done with these
parameters are labeled as ``fit/PSA'' in Tables \ref{tab:eightN} and 
\ref{tab:twentyN} and ``fit'' in the figures.

\begin{table}[pt]
\caption{Results for calculations of 8 neutron drops in harmonic potentials.
  All energies are in MeV and the rms radii $r_{\rm rms}$ are
  in fm. The NCFC results use up to the $N_{\rm max}$ in square brackets
  and parenthesis indicate the extrapolation uncertainty in the last quoted digit(s).  The approximations are explained in the text.}
  \label{tab:eightN}
  \smallskip
\begin{tabular}{ccccccc}
  \hline
 Approx. &  $\hbar\Omega$  & $E_{\rm tot}$ & $E_{\rm int}$ &
     $U_{\rm ext}$ & $r_{\rm rms}$ \\
  \hline
  \hline
  HF       & 5  & 71.9  & 37.6 & 34.3 & 3.78  \\
  HF/NV    & 5  & 72.3  & 37.5 & 34.8 & 3.80 \\
  HF/PSA    & 5  & 72.1  & 37.5 & 34.5 &3.78 \\
  BHF/PSA    & 5  & 68.8  & 34.9 & 33.9 & 3.75 \\
  fit/PSA    & 5  & 70.0  & 36.8 & 33.2 & 3.71 \\
  NCFC [14]    & 5  & $<69.5$  &  &  & \\
  \hline
  HF       & 10  & 142.4  & 69.6 & 72.8 &  2.75 \\
  HF/NV    & 10  &  144.5 & 69.5 & 75.0 &  2.79 \\
  HF/PSA    & 10  & 143.4  & 69.6 & 73.8 &  2.77 \\
  BHF/PSA    & 10  & 139.4  & 66.2 & 73.2 &  2.75 \\
  fit/PSA    & 10  & 138.6  & 67.3 & 71.3 &  2.72 \\
  NCFC [16]    & 10  & $138.1(6)$  & $66(2)$ & $72(2)$ &  $2.73(3)$ \\
  \hline
  HF       & 15  & 217.4  & 101.8 & 115.6 &  2.31 \\
  HF/NV    & 15  & 222.1  & 102.1 & 120.0 &  2.35 \\
  HF/PSA    & 15  & 219.5  & 101.9 & 117.6 &  2.33 \\
  BHF/PSA    & 15  & 214.8  & 98.5 & 116.2 &  2.31 \\
  fit/PSA    & 15  & 212.5  & 98.1 & 114.4 &  2.30 \\
  NCFC [16]    & 15  & $212.7(2)$  & $98.6(4)$  & $114.1(4)$ &  $2.293(4)$ \\
  \hline
  HF       & 20  & 296.4  & 135.1 & 161.3 &  2.04 \\
  HF/NV    & 20  & 304.1  & 136.3 & 167.8 &  2.09 \\
  HF/PSA    & 20  & 299.8  & 135.6 & 164.2 &  2.06 \\
  BHF/PSA    & 20  & 294.1  & 131.8 & 162.4 &  2.05   \\
  fit/PSA    & 20  & 290.9  & 130.0 & 160.9 & 2.04  \\
  NCFC [16]    & 20  & $290.8(2)$  & $131.5(3)$ & $159.3(3)$ &  $2.032(2)$ \\
  \hline
\end{tabular}
%\end{table}
%
\vspace*{.3in}
%
%\begin{table}
\caption{Results for calculations of 20 neutron drops in harmonic potentials
 with the same conventions as Table~\ref{tab:eightN}.}
  \label{tab:twentyN}
  \smallskip
\begin{tabular}{ccclll}
  \hline
 Approx. &  $\hbar\Omega$  & $E_{\rm tot}$ & $E_{\rm int}$ &
     $U_{\rm ext}$ & $r_{\rm rms}$ \\
  \hline
  \hline
  HF	  & 5  & 230.4  & 120.9 & 109.6 &  4.26 \\
  HF/NV   & 5  & 232.8  & 120.2 & 112.5 & 4.32  \\
  HF/PSA   & 5  & 231.3  & 120.7 & 110.6 & 4.28  \\
  BHF/PSA   & 5  & 221.3  & 112.4  & 108.9 & 4.25  \\
  fit/PSA   & 5  & 223.0  & 117.5 & 105.5 & 4.18  \\
  \hline
  HF      & 10  & 455.4  & 224.0 & 231.5 &  3.10 \\
  HF/NV   & 10  & 466.5  & 224.2 & 242.2 &  3.17 \\
  HF/PSA   & 10  & 459.3  & 224.1 & 235.2  &  3.12 \\
  BHF/PSA   & 10  & 445.0  & 215.7 & 229.3 &  3.08 \\
  fit/PSA   & 10  & 441.5  & 214.8 & 226.7 & 3.07  \\
  NCFC [8]   & 10  & ${<452.}$  &  &  &   \\
  \hline
  HF      & 15  & 693.0  & 328.1 & 364.9  & 2.59 \\
  HF/NV   & 15  & 715.7  & 332.5 & 383.2  & 2.66 \\
  HF/PSA   & 15  & 700.9  & 329.5 & 371.4  & 2.62 \\
  BHF/PSA   & 15  & 680.1  & 318.2 & 361.8  & 2.58 \\
  fit/PSA   & 15  & 675.9  & 313.2 & 362.7  & 2.59 \\
  NCFC [8]   & 15  &  $678(8)$ & $322(10)$ & $356(10)$  & $2.56(4)$ \\
  \hline
  HF      & 20  & 941.3  & 435.2 & 506.1  & 2.29 \\
  HF/NV   & 20  & 976.1  & 446.3 &  529.8 & 2.34 \\
  HF/PSA   & 20  & 953.7  & 438.4 & 515.4  & 2.31 \\
  BHF/PSA   & 20  & 928.1  & 417.7 & 510.4  & 2.30 \\
  fit/PSA   & 20  & 924.4  & 414.5 & 509.9  & 2.30 \\
  NCFC [8]   & 20  & $922(6)$  & $425(10)$  & $497(10)$  & $2.27(3)$ \\
  \hline
\end{tabular}
\end{table}

We have set the coupling constant $C^{\rho^2}_D$ to zero in the fit
to neutron matter properties.
In the usual Skyrme-DFT scheme, the density
dependence controlled by the power $\gamma$
is needed to produce reasonable saturation properties.
For example, the incompressibility of symmetric 
nuclear matter is strongly affected by $\gamma$, and usually an acceptable value requires $\gamma<1$.
In the present calculations we do not constrain symmetric matter;
indeed, the Minnesota potential does not produce realistic saturation.
More generally,
the density dependence from nonzero $C^{\rho^2}_D$
is used to effectively account for 
beyond-HF and three-body effects. 
The simplicity of the NN-only Minnesota potential seemingly
lets us transfer  beyond-HF effects to the other coupling constants
and omit the $C^{\rho^2}_D$ term entirely.

\begin{figure}[tbh]
 \includegraphics*[width=3.3in]{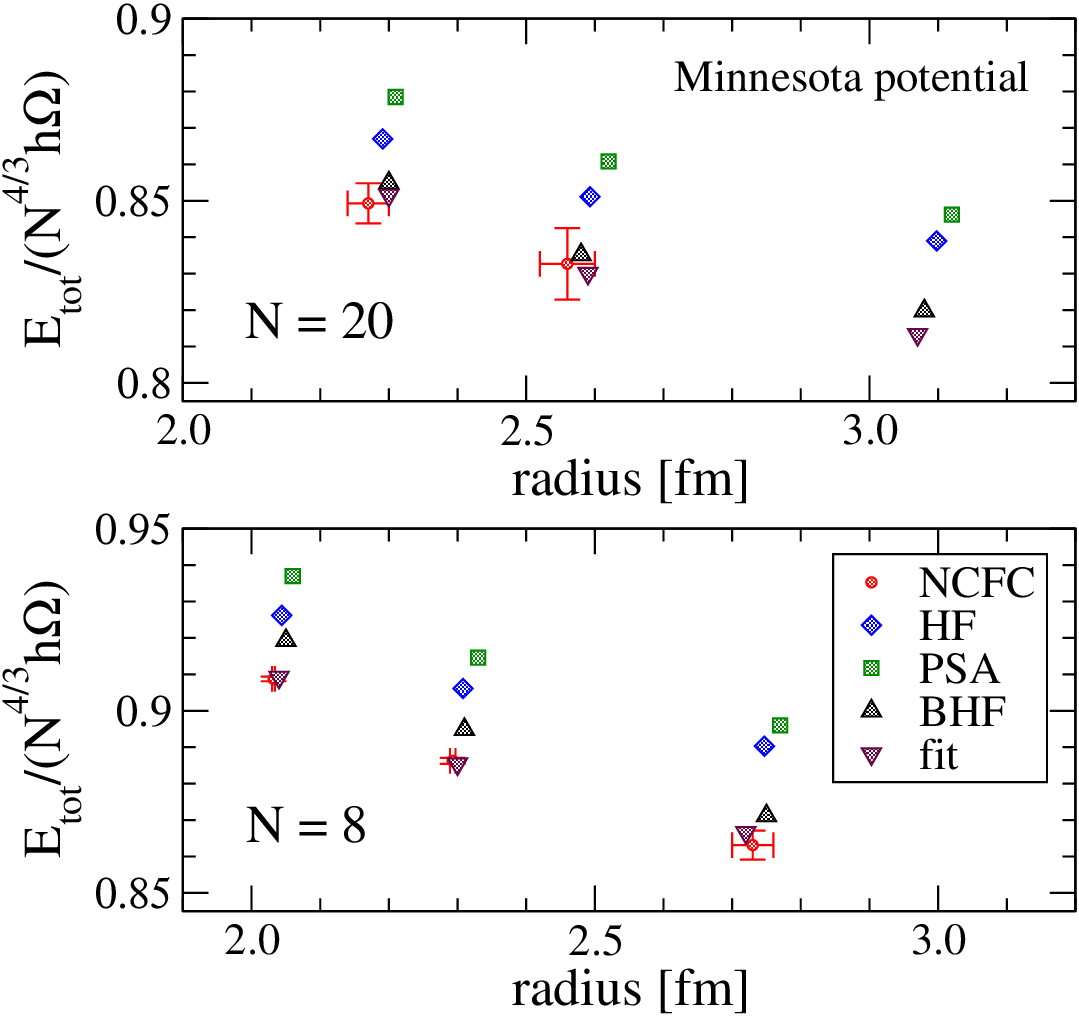}
 \caption{Total energy and radius
 for 8 and 20 neutrons in a harmonic
 potential with oscillator parameters 10\,MeV, 15\,MeV and 20\,MeV.  Calculations
 using the NCFC are compared to full HF, the
 DME/PSA approximation to HF with exact Hartree, 
 and results incorporating
 the resummed ladders for neutron matter using a density-dependent 
 adjustment (BHF) and using fit coefficients from Eq.~\eqref{eq:fitCset}.}
 \label{fig:energy_tot}
\end{figure}

\begin{figure}[h]
 \includegraphics*[width=3.3in]{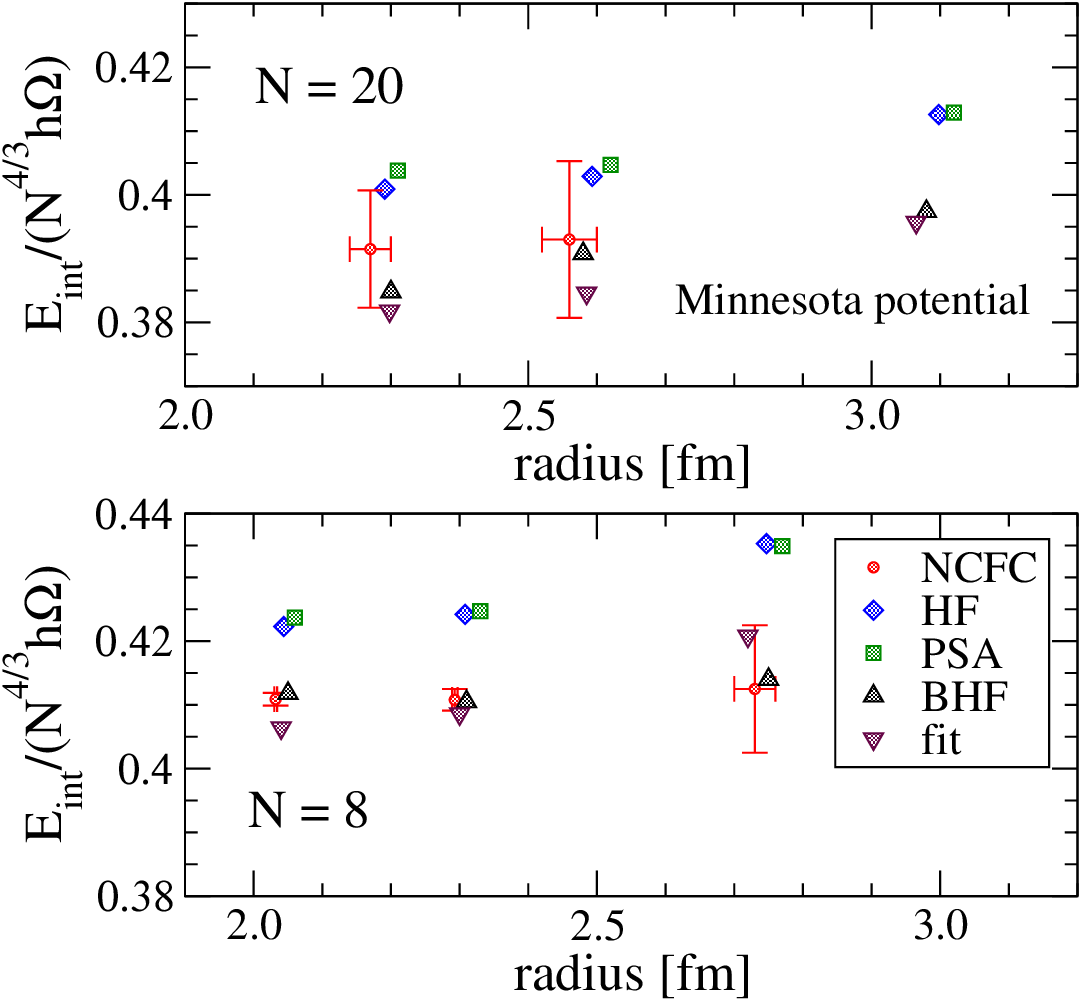}
 \caption{Internal energy and radius
 for 8 and 20 neutrons in a harmonic
 potential with oscillator parameters 10\,MeV, 15\,MeV and 20\,MeV
 as in Fig.~\ref{fig:energy_tot}.}
 \label{fig:energy_int}
\end{figure}

\begin{figure}[bht!]
 \includegraphics*[width=3.3in]{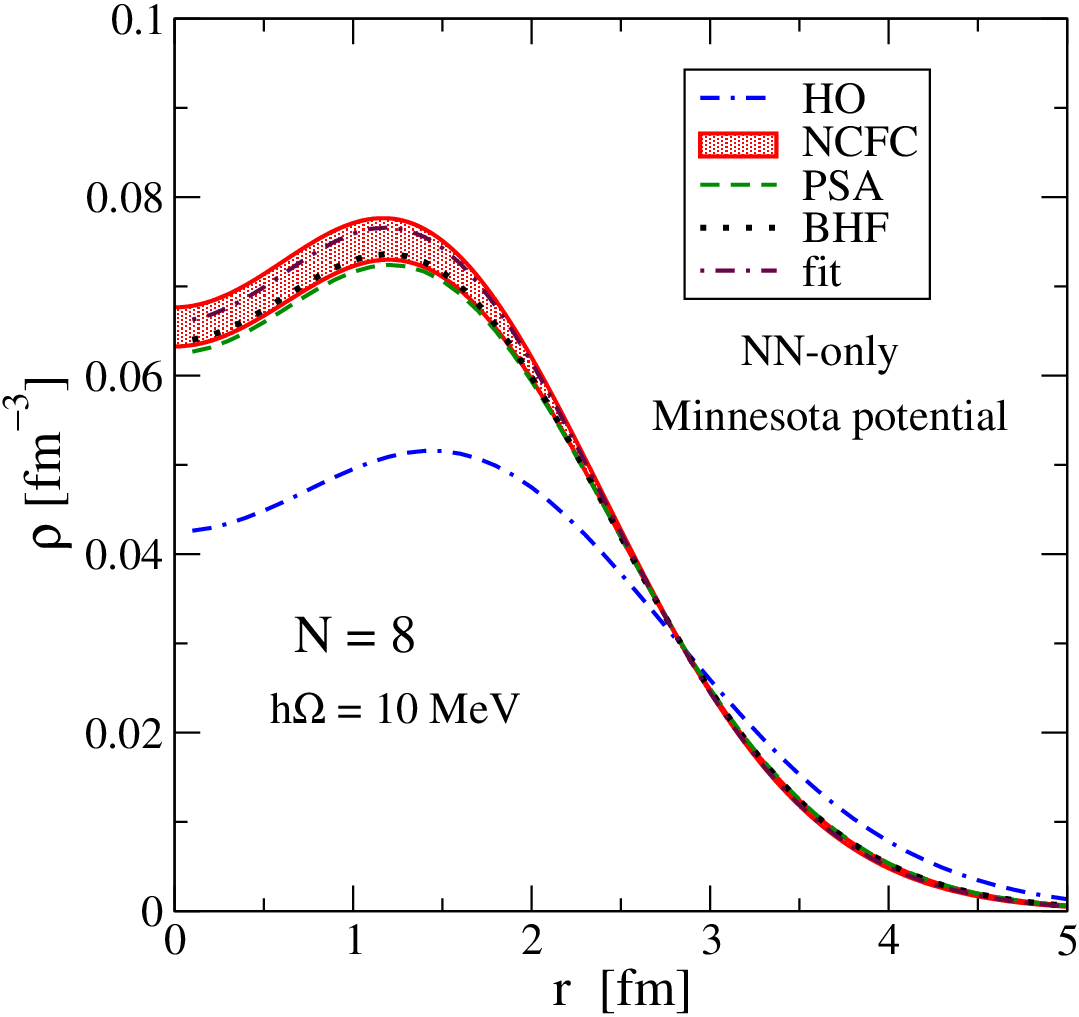}
 \caption{Density for 8 neutrons in a harmonic
 potential with oscillator parameter 10\,MeV.  The non-interacting density (HO) is compared to calculations
 using the NCFC (shaded region), the
 DME/PSA approximation to HF with exact Hartree, and results incorporating
 the resummed ladders for neutron matter via a density-dependent 
 adjustment (DME/PSA BHF) and using fit coefficients.}
 \label{fig:density}
\end{figure}

\begin{figure}[tbh]
 \includegraphics*[width=3.3in]{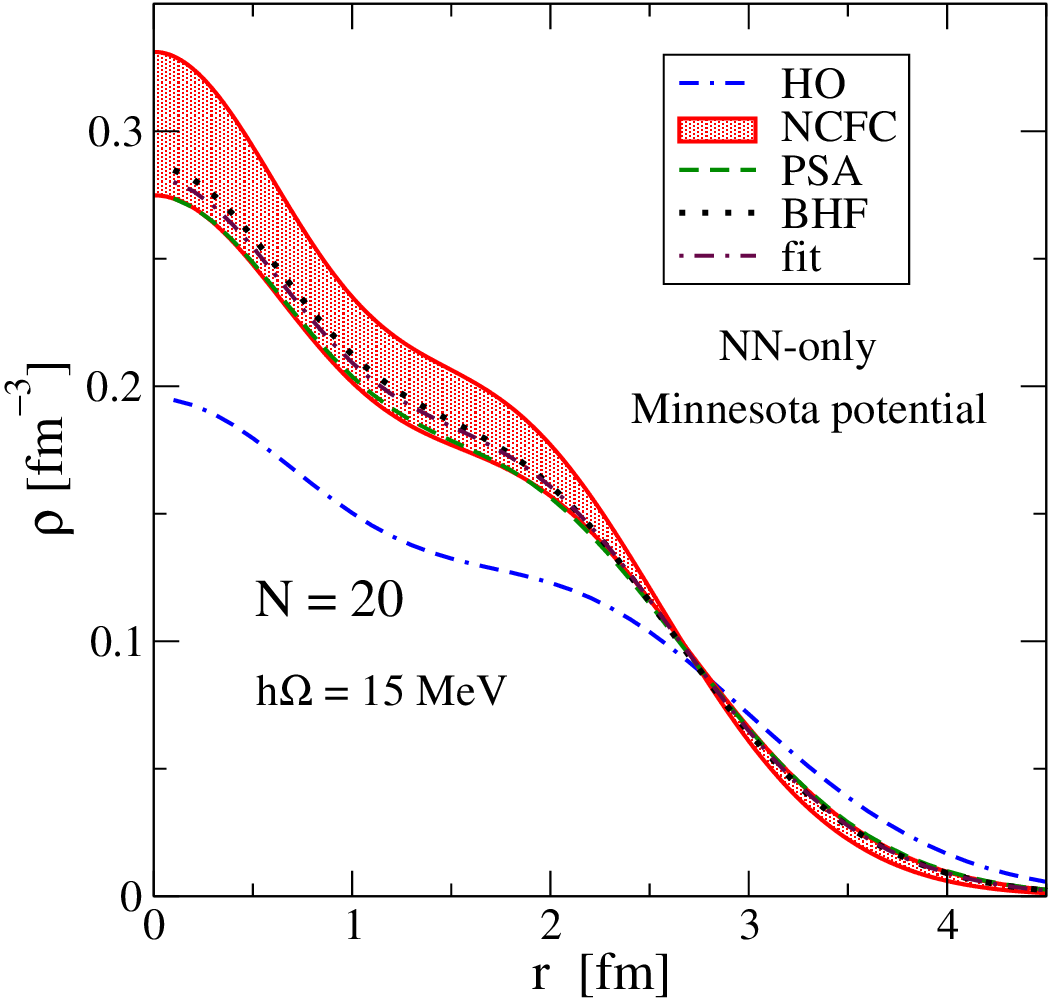}
 \caption{Density for 20 neutrons in a harmonic
 potential with oscillator parameter 15\,MeV, see
 Fig.~\ref{fig:density} caption.}
 \label{fig:density2}
\end{figure}

\begin{figure}[tbh]
 \includegraphics*[width=3.3in]{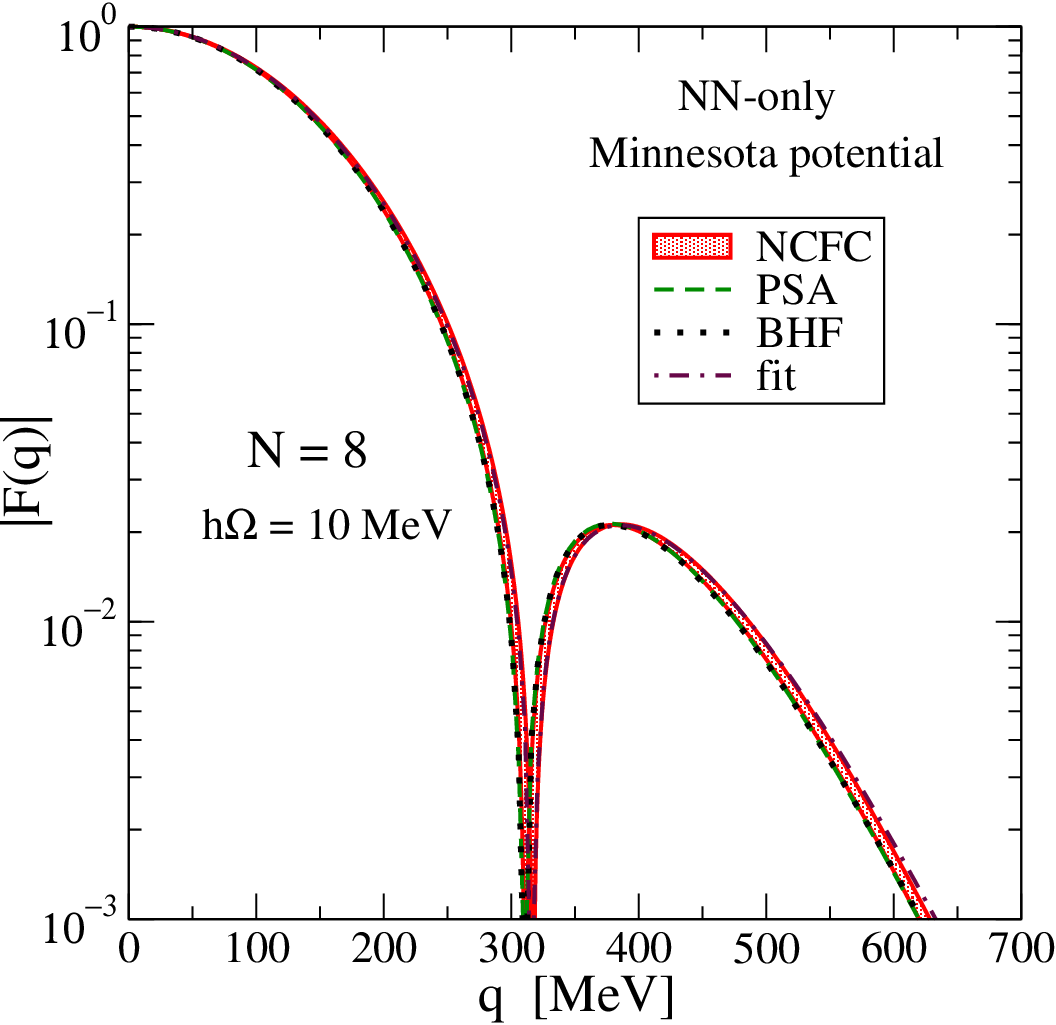}
 \caption{Form factor for 8 neutrons in a harmonic
 potential with oscillator parameter 10\,MeV, see
 Fig.~\ref{fig:density} caption.}
 \label{fig:formfactor}
\end{figure}

In Tables~\ref{tab:eightN} and \ref{tab:twentyN}, we summarize 
results for different trap parameters from the NV and PSA DME implementations
of HF with exact Hartree,
the two strategies to go beyond HF (BHF and fit, both based on PSA with
exact Hartree),
along with full HF and NCFC results. 
As already observed from the earlier tables,
the comparisons to HF shows that the DME-HF/NV and DME-HF/PSA calculations
have consistently higher energies and radii.
These trends and the comparison to the NCFC results are evident
in Figs.~\ref{fig:energy_tot} and \ref{fig:energy_int}, 
where energies and radii from the various DME prescriptions 
are compared to a full Hartree-Fock calculation and
to NCFC calculations for 8 and 20 neutrons.
The energies have been scaled by the Thomas-Fermi energy
trend $N^{4/3}\hbar\Omega$~\cite{Gandolfi:2010za} to remove
the dominant dependence on $N$ and $\hbar\Omega$.
The error bars from the NCFC are from the extrapolations;  the exact 
results for the Minnesota Hamiltonian are expected to lie within
these error ranges.

The DME/PSA BHF results improve on the HF and DME/PSA HF results
systematically for both the total and internal energies, and for the
radii.  They are within the errors of the NCFC in many cases.
The DME/PSA fit result with only
additional volume terms fit to neutron matter, 
is consistently worse
than the HF results, particularly for the internal energy  (not shown).
This failure is consistent with Ref.~\cite{Stoitsov:2010ha},
where the analogous prescription
was found to introduce unacceptably large over-binding in nuclei.
Here, the total energies and rms radii are systematically too small.
But once a surface term is added and fit to the NCFC results
for the total energies,
excellent systematics are found for these energies and 
the predicted radii.  The predicted internal energies are also
improved although there are much larger differences from the central
NCFC values than for the total energies.  

The reproduction of coordinate-space densities and the corresponding
momentum-space form factor are shown for some representative cases
in Figs.~\ref{fig:density}, \ref{fig:density2}, and \ref{fig:formfactor}.
These examples illustrate the very large range in interior density
probed by this set of external potentials.
The improvements noted for DMA/PSA BHF energies and radii 
are also seen for the
densities.
The DME/PSA fit results might be judged the best, 
although the error bands from the NCFC calculation are
too large to allow a definitive conclusion.
The form factor is given by:
\beq
   F(q) = \frac{1}{N} \int\! dr\, 4\pi r^2 \rho(r) \frac{\sin qr}{qr}
   \;.
\eeq
Not surprisingly, the characteristic features of the first minimum
and height of the second maximum are well reproduced within
the NCFC error band.

\section{Summary}
\label{sect:summary}

In this paper, we perform test calculations to help 
develop density functionals for nuclei using microscopic input.
In particular, we use the density matrix expansion (DME) as a bridge
from many-body perturbation theory to EDF's that can be
used in solvers and with optimization techniques 
developed for phenomenological Skyrme functionals.
There are many implementation questions and options for the DME, some of which
are addressed in the present work. 
Ultimately we will use high-precision two- and three-body nuclear 
interactions, such as from chiral effective field theory, evolved to 
softer forms using renormalization group methods.  This softening
makes them suitable for a MBPT treatment, unlike conventional interactions.
As an interim step, we have used the Minnesota potential as a
(moderately) soft, semi-realistic interaction for our tests. 
 
The test environment is interacting neutrons in a harmonic trap.
By varying the oscillator frequency of the trap, different inhomogeneous
density profiles are probed.  According to DFT, an EDF for self-bound
nuclei should be the same for the neutron systems, with the simple
addition of the external potential to the Kohn-Sham potential
and $U_{\rm ext}$ to the energy functional.
Thus we can make controlled explorations of the energies for different
density distributions. 
The predictions are validated against \abinitio\ calculations using
NCFC methods, made possible for 8 and 20 neutrons 
by computational and algorithmic advances enabled
by the UNEDF project.
 A key feature is that the same Hamiltonian is used for the
\abinitio\ solution and the functional.  This allows us to more reliably
isolate different sources of approximation errors.

Comparisons were first made between the DME at the HF level and full
HF calculations.  
The best results for the
improved form of the DME, which uses phase-space averaging (PSA),
were consistently superior to the original Negele-Vautherin formulation. 
This agrees with the
findings in Ref.~\cite{Carlsson:2010da}, where it was found in non-self-consistent
calculations of nuclei that the PSA-DME was the most accurate formulation when the expansion
is truncated at second order in gradients.
(Note: the improvement will be more significant
for more realistic interactions including spin-orbit and tensor
forces.)  Several options for treating the Hartree
(direct) contribution were considered, with a clear preference for
an exact treatment in the PSA version of DME.
This is consistent with past experience applying the DME.

There are still systematic discrepancies between the
best DME Hartree-Fock calculations and the full HF results.
This is in contrast to results from recent exact-exchange orbital-based
calculations, which find extremely close agreement with HF for the
same neutron systems and interactions~\cite{Drut:2011te}.
Thus the source of the discrepancies seems to be the DME approximation 
and not simply the use of a local Kohn-Sham potential. 
This implies that DME-based functionals will always need to be
supplemented with some correction mechanism for inherent DME approximations
as well as for errors from many-body approximations.

Hartree-Fock is a good quantitative starting point for these neutron drop
systems with the Minnesota potential, but the full NCFC solutions
show clear differences in the patterns of energies and radii.
These differences are used as a testbed for two ways
to incorporate correlations beyond HF in the DME functional.
In both cases, Brueckner-Hartree-Fock (BHF) calculations of neutron
matter with the Minnesota potential were used as the exact reference.
(Instead, one could use \abinitio\ calculations if available, or else
calibrate to phenomenological values.)

In the first approach, the functional was modified by density-dependent
terms, while in the second approach, the interaction was supplemented
with Skyrme-like contact terms, whose coefficients were fit to
the neutron matter calculation.
The first method (labeled BHF in the figures), showed systematic
improvement from DME Hartree-Fock toward the full NCFC results, both for
the energies and the radius.
The second method with volume terms only was found to be unacceptable, 
but the addition of a fit surface term 
improved the systematics dramatically,
to roughly the error limits of the current NCFC calculations
for the total energies and the predicted radii.
These results validate the strategies planned for more
realistic forces.

There are many ways forward from the present calculations.
The neutron drop system will continue to be a valuable tool for
diagnostic testing, which will include using non-harmonic (e.g.,
Woods-Saxon) and
deformed traps.
On-going development of the density matrix expansion includes
the extension to pairing, which will be tested using open-shell
neutron numbers, and extensions to incorporate
higher-order MBPT and more complete
low-momentum potentials.
These extensions
will be tested both in the neutron drop systems and for ordinary self-bound
nuclei.
Work in these directions is in progress.

\medskip

\section*{Acknowledgments}
 
We thank J. Drut and L. Platter for useful discussions. This work was supported by the Office of Nuclear Physics, U.S. Department 
of Energy under Contract Nos. DE-FC02-09ER41583 and DE-FC02-09ER41582 (UNEDF SciDAC Collaboration), 
DE-FG02-96ER40963 and DE-FG02-07ER41529 (University of Tennessee),  DE-FG0587ER40361 (Joint Institute 
for Heavy Ion Research), DE-FC02-09ER41581, DE-FG02-87ER40371 and DE-FC02-09ER41585. Computational resources were provided through an INCITE award ``Computational 
Nuclear Structure" by the National Center for Computational Sciences (NCCS) and 
the National Institute for Computational Sciences (NICS) at Oak Ridge National Laboratory. Additional computational resources were provided by the National Energy Research Scientific Computing Center (NERSC) at Lawrence Berkeley National Laboratory.

%\begin{references}
%\end{references}

\bibliographystyle{apsrev}
\bibliography{vlowk_refs}

\begin{thebibliography}{46}
\expandafter\ifx\csname natexlab\endcsname\relax\def\natexlab#1{#1}\fi
\expandafter\ifx\csname bibnamefont\endcsname\relax
  \def\bibnamefont#1{#1}\fi
\expandafter\ifx\csname bibfnamefont\endcsname\relax
  \def\bibfnamefont#1{#1}\fi
\expandafter\ifx\csname citenamefont\endcsname\relax
  \def\citenamefont#1{#1}\fi
\expandafter\ifx\csname url\endcsname\relax
  \def\url#1{\texttt{#1}}\fi
\expandafter\ifx\csname urlprefix\endcsname\relax\def\urlprefix{URL }\fi
\providecommand{\bibinfo}[2]{#2}
\providecommand{\eprint}[2][]{\url{#2}}

\bibitem[{\citenamefont{Pieper}(2005)}]{Pieper:2004qh}
\bibinfo{author}{\bibfnamefont{S.~C.} \bibnamefont{Pieper}},
  \bibinfo{journal}{Nucl. Phys. A} \textbf{\bibinfo{volume}{751}},
  \bibinfo{pages}{516} (\bibinfo{year}{2005}), \eprint{nucl-th/0410115}.

\bibitem[{\citenamefont{Quaglioni and Navratil}(2008)}]{Quaglioni:2007qe}
\bibinfo{author}{\bibfnamefont{S.}~\bibnamefont{Quaglioni}} \bibnamefont{and}
  \bibinfo{author}{\bibfnamefont{P.}~\bibnamefont{Navratil}},
  \bibinfo{journal}{Few Body Syst.} \textbf{\bibinfo{volume}{44}},
  \bibinfo{pages}{337} (\bibinfo{year}{2008}), \eprint{0712.0855}.

\bibitem[{\citenamefont{Hagen et~al.}(2008)\citenamefont{Hagen, Papenbrock,
  Dean, and Hjorth-Jensen}}]{Hagen:2008iw}
\bibinfo{author}{\bibfnamefont{G.}~\bibnamefont{Hagen}},
  \bibinfo{author}{\bibfnamefont{T.}~\bibnamefont{Papenbrock}},
  \bibinfo{author}{\bibfnamefont{D.~J.} \bibnamefont{Dean}}, \bibnamefont{and}
  \bibinfo{author}{\bibfnamefont{M.}~\bibnamefont{Hjorth-Jensen}},
  \bibinfo{journal}{Phys. Rev. Lett.} \textbf{\bibinfo{volume}{101}},
  \bibinfo{pages}{092502} (\bibinfo{year}{2008}), \eprint{0806.3478}.

\bibitem[{\citenamefont{Maris et~al.}(2010{\natexlab{a}})\citenamefont{Maris,
  Shirokov, and Vary}}]{Maris:2009bx}
\bibinfo{author}{\bibfnamefont{P.}~\bibnamefont{Maris}},
  \bibinfo{author}{\bibfnamefont{A.~M.} \bibnamefont{Shirokov}},
  \bibnamefont{and} \bibinfo{author}{\bibfnamefont{J.~P.} \bibnamefont{Vary}},
  \bibinfo{journal}{Phys. Rev.} \textbf{\bibinfo{volume}{C81}},
  \bibinfo{pages}{021301} (\bibinfo{year}{2010}{\natexlab{a}}),
  \eprint{0911.2281}.

\bibitem[{\citenamefont{Bender et~al.}(2003)\citenamefont{Bender, Heenen, and
  Reinhard}}]{Bender:2003jk}
\bibinfo{author}{\bibfnamefont{M.}~\bibnamefont{Bender}},
  \bibinfo{author}{\bibfnamefont{P.-H.} \bibnamefont{Heenen}},
  \bibnamefont{and} \bibinfo{author}{\bibfnamefont{P.-G.}
  \bibnamefont{Reinhard}}, \bibinfo{journal}{Rev. Mod. Phys.}
  \textbf{\bibinfo{volume}{75}}, \bibinfo{pages}{121} (\bibinfo{year}{2003}).

\bibitem[{\citenamefont{Maris et~al.}(2009)\citenamefont{Maris, Vary, and
  Shirokov}}]{Maris:2008ax}
\bibinfo{author}{\bibfnamefont{P.}~\bibnamefont{Maris}},
  \bibinfo{author}{\bibfnamefont{J.~P.} \bibnamefont{Vary}}, \bibnamefont{and}
  \bibinfo{author}{\bibfnamefont{A.~M.} \bibnamefont{Shirokov}},
  \bibinfo{journal}{Phys. Rev. C} \textbf{\bibinfo{volume}{79}},
  \bibinfo{pages}{014308} (\bibinfo{year}{2009}), \eprint{0808.3420}.

\bibitem[{\citenamefont{Bertsch et~al.}(2007)\citenamefont{Bertsch, Dean, and
  Nazarewicz}}]{unedf:2007}
\bibinfo{author}{\bibfnamefont{G.~F.} \bibnamefont{Bertsch}},
  \bibinfo{author}{\bibfnamefont{D.~J.} \bibnamefont{Dean}}, \bibnamefont{and}
  \bibinfo{author}{\bibfnamefont{W.}~\bibnamefont{Nazarewicz}},
  \bibinfo{journal}{SciDAC Review} \textbf{\bibinfo{volume}{6}},
  \bibinfo{pages}{42} (\bibinfo{year}{2007}).

\bibitem[{une()}]{unedfweb}
\bibinfo{howpublished}{For more details, see: \url{http://unedf.org}}.

\bibitem[{\citenamefont{Bogner et~al.}(2009)\citenamefont{Bogner, Furnstahl,
  and Platter}}]{Bogner:2008kj}
\bibinfo{author}{\bibfnamefont{S.~K.} \bibnamefont{Bogner}},
  \bibinfo{author}{\bibfnamefont{R.~J.} \bibnamefont{Furnstahl}},
  \bibnamefont{and} \bibinfo{author}{\bibfnamefont{L.}~\bibnamefont{Platter}},
  \bibinfo{journal}{Eur. Phys. J. A} \textbf{\bibinfo{volume}{39}},
  \bibinfo{pages}{219} (\bibinfo{year}{2009}), \eprint{0811.4198}.

\bibitem[{\citenamefont{Drut et~al.}(2010)\citenamefont{Drut, Furnstahl, and
  Platter}}]{Drut:2009ce}
\bibinfo{author}{\bibfnamefont{J.~E.} \bibnamefont{Drut}},
  \bibinfo{author}{\bibfnamefont{R.~J.} \bibnamefont{Furnstahl}},
  \bibnamefont{and} \bibinfo{author}{\bibfnamefont{L.}~\bibnamefont{Platter}},
  \bibinfo{journal}{Prog. Part. Nucl. Phys.} \textbf{\bibinfo{volume}{64}},
  \bibinfo{pages}{120} (\bibinfo{year}{2010}), \eprint{0906.1463}.

\bibitem[{\citenamefont{Gebremariam et~al.}(2010)\citenamefont{Gebremariam,
  Duguet, and Bogner}}]{Gebremariam:2009ff}
\bibinfo{author}{\bibfnamefont{B.}~\bibnamefont{Gebremariam}},
  \bibinfo{author}{\bibfnamefont{T.}~\bibnamefont{Duguet}}, \bibnamefont{and}
  \bibinfo{author}{\bibfnamefont{S.~K.} \bibnamefont{Bogner}},
  \bibinfo{journal}{Phys. Rev.} \textbf{\bibinfo{volume}{C82}},
  \bibinfo{pages}{014305} (\bibinfo{year}{2010}), \eprint{0910.4979}.

\bibitem[{\citenamefont{Gebremariam et~al.}(2011)\citenamefont{Gebremariam,
  Bogner, and Duguet}}]{Gebremariam:2010ni}
\bibinfo{author}{\bibfnamefont{B.}~\bibnamefont{Gebremariam}},
  \bibinfo{author}{\bibfnamefont{S.~K.} \bibnamefont{Bogner}},
  \bibnamefont{and} \bibinfo{author}{\bibfnamefont{T.}~\bibnamefont{Duguet}},
  \bibinfo{journal}{Nucl. Phys.} \textbf{\bibinfo{volume}{A851}},
  \bibinfo{pages}{17} (\bibinfo{year}{2011}), \eprint{1003.5210}.

\bibitem[{\citenamefont{Stoitsov et~al.}(2010)}]{Stoitsov:2010ha}
\bibinfo{author}{\bibfnamefont{M.}~\bibnamefont{Stoitsov}}
  \bibnamefont{et~al.}, \bibinfo{journal}{Phys. Rev.}
  \textbf{\bibinfo{volume}{C82}}, \bibinfo{pages}{054307}
  (\bibinfo{year}{2010}), \eprint{1009.3452}.

\bibitem[{\citenamefont{Kortelainen et~al.}(2010)}]{Kortelainen:2010hv}
\bibinfo{author}{\bibfnamefont{M.}~\bibnamefont{Kortelainen}}
  \bibnamefont{et~al.}, \bibinfo{journal}{Phys. Rev.}
  \textbf{\bibinfo{volume}{C82}}, \bibinfo{pages}{024313}
  (\bibinfo{year}{2010}), \eprint{1005.5145}.

\bibitem[{\citenamefont{Maris et~al.}(2010{\natexlab{b}})\citenamefont{Maris,
  Sosonkina, Vary, Ng, and Yang}}]{Maris:2010aaa}
\bibinfo{author}{\bibfnamefont{P.}~\bibnamefont{Maris}},
  \bibinfo{author}{\bibfnamefont{M.}~\bibnamefont{Sosonkina}},
  \bibinfo{author}{\bibfnamefont{J.~P.} \bibnamefont{Vary}},
  \bibinfo{author}{\bibfnamefont{E.~G.} \bibnamefont{Ng}}, \bibnamefont{and}
  \bibinfo{author}{\bibfnamefont{C.}~\bibnamefont{Yang}},
  \bibinfo{journal}{Proc. Comp. Sci.} \textbf{\bibinfo{volume}{1}},
  \bibinfo{pages}{97} (\bibinfo{year}{2010}{\natexlab{b}}).

\bibitem[{\citenamefont{Chabanat et~al.}(1997)\citenamefont{Chabanat, Meyer,
  Bonche, Schaeffer, and Haensel}}]{Chabanat:1997qh}
\bibinfo{author}{\bibfnamefont{E.}~\bibnamefont{Chabanat}},
  \bibinfo{author}{\bibfnamefont{J.}~\bibnamefont{Meyer}},
  \bibinfo{author}{\bibfnamefont{P.}~\bibnamefont{Bonche}},
  \bibinfo{author}{\bibfnamefont{R.}~\bibnamefont{Schaeffer}},
  \bibnamefont{and} \bibinfo{author}{\bibfnamefont{P.}~\bibnamefont{Haensel}},
  \bibinfo{journal}{Nucl. Phys.} \textbf{\bibinfo{volume}{A627}},
  \bibinfo{pages}{710} (\bibinfo{year}{1997}).

\bibitem[{\citenamefont{Alex~Brown}(1998)}]{AlexBrown:1998zz}
\bibinfo{author}{\bibfnamefont{B.}~\bibnamefont{Alex~Brown}},
  \bibinfo{journal}{Phys. Rev.} \textbf{\bibinfo{volume}{C58}},
  \bibinfo{pages}{220} (\bibinfo{year}{1998}).

\bibitem[{\citenamefont{Gandolfi et~al.}(2011)\citenamefont{Gandolfi, Carlson,
  and Pieper}}]{Gandolfi:2010za}
\bibinfo{author}{\bibfnamefont{S.}~\bibnamefont{Gandolfi}},
  \bibinfo{author}{\bibfnamefont{J.}~\bibnamefont{Carlson}}, \bibnamefont{and}
  \bibinfo{author}{\bibfnamefont{S.~C.} \bibnamefont{Pieper}},
  \bibinfo{journal}{Phys. Rev. Lett.} \textbf{\bibinfo{volume}{106}},
  \bibinfo{pages}{012501} (\bibinfo{year}{2011}), \eprint{1010.4583}.

\bibitem[{\citenamefont{Carlson et~al.}(2011)\citenamefont{Carlson, Gandolfi,
  Maris, Vary, and Pieper}}]{Carlson:2011aa}
\bibinfo{author}{\bibfnamefont{J.}~\bibnamefont{Carlson}},
  \bibinfo{author}{\bibfnamefont{S.}~\bibnamefont{Gandolfi}},
  \bibinfo{author}{\bibfnamefont{P.}~\bibnamefont{Maris}},
  \bibinfo{author}{\bibfnamefont{J.}~\bibnamefont{Vary}}, \bibnamefont{and}
  \bibinfo{author}{\bibfnamefont{S.}~\bibnamefont{Pieper}}
  (\bibinfo{year}{2011}), \bibinfo{note}{in preparation}.

\bibitem[{\citenamefont{Duguet and Sadoudi}(2010)}]{Duguet:2010cv}
\bibinfo{author}{\bibfnamefont{T.}~\bibnamefont{Duguet}} \bibnamefont{and}
  \bibinfo{author}{\bibfnamefont{J.}~\bibnamefont{Sadoudi}},
  \bibinfo{journal}{J. Phys.} \textbf{\bibinfo{volume}{G37}},
  \bibinfo{pages}{064009} (\bibinfo{year}{2010}), \eprint{1001.0673}.

\bibitem[{\citenamefont{Negele and Vautherin}(1972)}]{Negele:1972zp}
\bibinfo{author}{\bibfnamefont{J.~W.} \bibnamefont{Negele}} \bibnamefont{and}
  \bibinfo{author}{\bibfnamefont{D.}~\bibnamefont{Vautherin}},
  \bibinfo{journal}{Phys. Rev. C} \textbf{\bibinfo{volume}{5}},
  \bibinfo{pages}{1472} (\bibinfo{year}{1972}).

\bibitem[{\citenamefont{Negele and Vautherin}(1975)}]{Negele:1975zz}
\bibinfo{author}{\bibfnamefont{J.~W.} \bibnamefont{Negele}} \bibnamefont{and}
  \bibinfo{author}{\bibfnamefont{D.}~\bibnamefont{Vautherin}},
  \bibinfo{journal}{Phys. Rev. C} \textbf{\bibinfo{volume}{11}},
  \bibinfo{pages}{1031} (\bibinfo{year}{1975}).

\bibitem[{\citenamefont{{Durand} et~al.}(1982)\citenamefont{{Durand},
  {Ramamurthy}, and {Schuck}}}]{Durand:1982aa}
\bibinfo{author}{\bibfnamefont{M.}~\bibnamefont{{Durand}}},
  \bibinfo{author}{\bibfnamefont{V.~S.} \bibnamefont{{Ramamurthy}}},
  \bibnamefont{and} \bibinfo{author}{\bibfnamefont{P.}~\bibnamefont{{Schuck}}},
  \bibinfo{journal}{Phys. Lett. B} \textbf{\bibinfo{volume}{113}},
  \bibinfo{pages}{116} (\bibinfo{year}{1982}).

\bibitem[{\citenamefont{Bulgac and Thompson}(1996)}]{Bulgac:1996aa}
\bibinfo{author}{\bibfnamefont{A.}~\bibnamefont{Bulgac}} \bibnamefont{and}
  \bibinfo{author}{\bibfnamefont{J.~M.} \bibnamefont{Thompson}},
  \bibinfo{journal}{Phys. Lett. B} \textbf{\bibinfo{volume}{383}},
  \bibinfo{pages}{127} (\bibinfo{year}{1996}).

\bibitem[{\citenamefont{Campi and Bouyssy}(1978)}]{Campi:1978}
\bibinfo{author}{\bibfnamefont{X.}~\bibnamefont{Campi}} \bibnamefont{and}
  \bibinfo{author}{\bibfnamefont{A.}~\bibnamefont{Bouyssy}},
  \bibinfo{journal}{Phys. Lett. B} \textbf{\bibinfo{volume}{73}},
  \bibinfo{pages}{263} (\bibinfo{year}{1978}).

\bibitem[{\citenamefont{Sprung et~al.}(1975)\citenamefont{Sprung, Vallieres,
  Campi, and Ko}}]{Sprung:1975}
\bibinfo{author}{\bibfnamefont{D.}~\bibnamefont{Sprung}},
  \bibinfo{author}{\bibfnamefont{M.}~\bibnamefont{Vallieres}},
  \bibinfo{author}{\bibfnamefont{X.}~\bibnamefont{Campi}}, \bibnamefont{and}
  \bibinfo{author}{\bibfnamefont{C.-M.} \bibnamefont{Ko}},
  \bibinfo{journal}{Nucl. Phys. A} \textbf{\bibinfo{volume}{253}},
  \bibinfo{pages}{1} (\bibinfo{year}{1975}).

\bibitem[{\citenamefont{Carlsson and Dobaczewski}(2010)}]{Carlsson:2010da}
\bibinfo{author}{\bibfnamefont{B.~G.} \bibnamefont{Carlsson}} \bibnamefont{and}
  \bibinfo{author}{\bibfnamefont{J.}~\bibnamefont{Dobaczewski}},
  \bibinfo{journal}{Phys. Rev. Lett.} \textbf{\bibinfo{volume}{105}},
  \bibinfo{pages}{122501} (\bibinfo{year}{2010}), \eprint{1003.2543}.

\bibitem[{\citenamefont{Hofmann and Lenske}(1998)}]{Hofmann:1997zu}
\bibinfo{author}{\bibfnamefont{F.}~\bibnamefont{Hofmann}} \bibnamefont{and}
  \bibinfo{author}{\bibfnamefont{H.}~\bibnamefont{Lenske}},
  \bibinfo{journal}{Phys. Rev.} \textbf{\bibinfo{volume}{C57}},
  \bibinfo{pages}{2281} (\bibinfo{year}{1998}), \eprint{nucl-th/9705049}.

\bibitem[{\citenamefont{Kaiser et~al.}(2003)\citenamefont{Kaiser, Fritsch, and
  Weise}}]{Kaiser:2002jz}
\bibinfo{author}{\bibfnamefont{N.}~\bibnamefont{Kaiser}},
  \bibinfo{author}{\bibfnamefont{S.}~\bibnamefont{Fritsch}}, \bibnamefont{and}
  \bibinfo{author}{\bibfnamefont{W.}~\bibnamefont{Weise}},
  \bibinfo{journal}{Nucl. Phys. A} \textbf{\bibinfo{volume}{724}},
  \bibinfo{pages}{47} (\bibinfo{year}{2003}), \eprint{nucl-th/0212049}.

\bibitem[{\citenamefont{Kaiser and Weise}(2010)}]{Kaiser:2009me}
\bibinfo{author}{\bibfnamefont{N.}~\bibnamefont{Kaiser}} \bibnamefont{and}
  \bibinfo{author}{\bibfnamefont{W.}~\bibnamefont{Weise}},
  \bibinfo{journal}{Nucl. Phys.} \textbf{\bibinfo{volume}{A836}},
  \bibinfo{pages}{256} (\bibinfo{year}{2010}), \eprint{0912.3207}.

\bibitem[{\citenamefont{Thompson et~al.}(1977)\citenamefont{Thompson, Lemere,
  and Tang}}]{Thompson:1977zz}
\bibinfo{author}{\bibfnamefont{D.~R.} \bibnamefont{Thompson}},
  \bibinfo{author}{\bibfnamefont{M.}~\bibnamefont{Lemere}}, \bibnamefont{and}
  \bibinfo{author}{\bibfnamefont{Y.~C.} \bibnamefont{Tang}},
  \bibinfo{journal}{Nucl. Phys.} \textbf{\bibinfo{volume}{A286}},
  \bibinfo{pages}{53} (\bibinfo{year}{1977}).

\bibitem[{\citenamefont{Bennaceur and Dobaczewski}(2005)}]{Bennaceur:2005mx}
\bibinfo{author}{\bibfnamefont{K.}~\bibnamefont{Bennaceur}} \bibnamefont{and}
  \bibinfo{author}{\bibfnamefont{J.}~\bibnamefont{Dobaczewski}},
  \bibinfo{journal}{Comput. Phys. Commun.} \textbf{\bibinfo{volume}{168}},
  \bibinfo{pages}{96} (\bibinfo{year}{2005}), \eprint{nucl-th/0501002}.

\bibitem[{\citenamefont{Stoitsov et~al.}(2005)\citenamefont{Stoitsov,
  Dobaczewski, Nazarewicz, and Ring}}]{Stoitsov:2004pe}
\bibinfo{author}{\bibfnamefont{M.~V.} \bibnamefont{Stoitsov}},
  \bibinfo{author}{\bibfnamefont{J.}~\bibnamefont{Dobaczewski}},
  \bibinfo{author}{\bibfnamefont{W.}~\bibnamefont{Nazarewicz}},
  \bibnamefont{and} \bibinfo{author}{\bibfnamefont{P.}~\bibnamefont{Ring}},
  \bibinfo{journal}{Comput. Phys. Commun.} \textbf{\bibinfo{volume}{167}},
  \bibinfo{pages}{43} (\bibinfo{year}{2005}), \eprint{nucl-th/0406075}.

\bibitem[{\citenamefont{Kortelainen and Stoitsov}(2011)}]{dmemodule}
\bibinfo{author}{\bibfnamefont{M.}~\bibnamefont{Kortelainen}} \bibnamefont{and}
  \bibinfo{author}{\bibfnamefont{M.}~\bibnamefont{Stoitsov}}
  (\bibinfo{year}{2011}), \bibinfo{note}{to be published}.

\bibitem[{\citenamefont{Bogner et~al.}(2008)}]{Bogner:2007rx}
\bibinfo{author}{\bibfnamefont{S.~K.} \bibnamefont{Bogner}}
  \bibnamefont{et~al.}, \bibinfo{journal}{Nucl. Phys. A}
  \textbf{\bibinfo{volume}{801}}, \bibinfo{pages}{21} (\bibinfo{year}{2008}),
  \eprint{0708.3754}.

\bibitem[{\citenamefont{Zheng et~al.}(1994)\citenamefont{Zheng, Vary, and
  Barrett}}]{Zheng:1994jr}
\bibinfo{author}{\bibfnamefont{D.~C.} \bibnamefont{Zheng}},
  \bibinfo{author}{\bibfnamefont{J.~P.} \bibnamefont{Vary}}, \bibnamefont{and}
  \bibinfo{author}{\bibfnamefont{B.~R.} \bibnamefont{Barrett}},
  \bibinfo{journal}{Phys. Rev.} \textbf{\bibinfo{volume}{C50}},
  \bibinfo{pages}{2841} (\bibinfo{year}{1994}), \eprint{nucl-th/9405018}.

\bibitem[{\citenamefont{Zheng et~al.}(1995)\citenamefont{Zheng, Barrett, Vary,
  Haxton, and Song}}]{Zheng:1995td}
\bibinfo{author}{\bibfnamefont{D.~C.} \bibnamefont{Zheng}},
  \bibinfo{author}{\bibfnamefont{B.~R.} \bibnamefont{Barrett}},
  \bibinfo{author}{\bibfnamefont{J.~P.} \bibnamefont{Vary}},
  \bibinfo{author}{\bibfnamefont{W.~C.} \bibnamefont{Haxton}},
  \bibnamefont{and} \bibinfo{author}{\bibfnamefont{C.~L.} \bibnamefont{Song}},
  \bibinfo{journal}{Phys. Rev.} \textbf{\bibinfo{volume}{C52}},
  \bibinfo{pages}{2488} (\bibinfo{year}{1995}).

\bibitem[{\citenamefont{Navratil
  et~al.}(2000{\natexlab{a}})\citenamefont{Navratil, Vary, and
  Barrett}}]{Navratil:2000ww}
\bibinfo{author}{\bibfnamefont{P.}~\bibnamefont{Navratil}},
  \bibinfo{author}{\bibfnamefont{J.~P.} \bibnamefont{Vary}}, \bibnamefont{and}
  \bibinfo{author}{\bibfnamefont{B.~R.} \bibnamefont{Barrett}},
  \bibinfo{journal}{Phys. Rev. Lett.} \textbf{\bibinfo{volume}{84}},
  \bibinfo{pages}{5728} (\bibinfo{year}{2000}{\natexlab{a}}),
  \eprint{nucl-th/0004058}.

\bibitem[{\citenamefont{Navratil
  et~al.}(2000{\natexlab{b}})\citenamefont{Navratil, Vary, and
  Barrett}}]{Navratil:2000gs}
\bibinfo{author}{\bibfnamefont{P.}~\bibnamefont{Navratil}},
  \bibinfo{author}{\bibfnamefont{J.~P.} \bibnamefont{Vary}}, \bibnamefont{and}
  \bibinfo{author}{\bibfnamefont{B.~R.} \bibnamefont{Barrett}},
  \bibinfo{journal}{Phys. Rev. C} \textbf{\bibinfo{volume}{62}},
  \bibinfo{pages}{054311} (\bibinfo{year}{2000}{\natexlab{b}}).

\bibitem[{\citenamefont{Vary}(1992 (unpublished))}]{Vary:1992mfd}
\bibinfo{author}{\bibfnamefont{J.}~\bibnamefont{Vary}},
  \emph{\bibinfo{title}{The Many-Fermion-Dynamics Shell-Model Code}}
  (\bibinfo{publisher}{Iowa Sate University}, \bibinfo{year}{1992
  (unpublished)}).

\bibitem[{\citenamefont{Sternberg et~al.}(2008)\citenamefont{Sternberg, Ng,
  Yang, Maris, Vary, Sosonkina, and Le}}]{Sternberg:2008aa}
\bibinfo{author}{\bibfnamefont{P.}~\bibnamefont{Sternberg}},
  \bibinfo{author}{\bibfnamefont{E.~G.} \bibnamefont{Ng}},
  \bibinfo{author}{\bibfnamefont{C.}~\bibnamefont{Yang}},
  \bibinfo{author}{\bibfnamefont{P.}~\bibnamefont{Maris}},
  \bibinfo{author}{\bibfnamefont{J.~P.} \bibnamefont{Vary}},
  \bibinfo{author}{\bibfnamefont{M.}~\bibnamefont{Sosonkina}},
  \bibnamefont{and} \bibinfo{author}{\bibfnamefont{H.~V.} \bibnamefont{Le}}
  (\bibinfo{year}{2008}), \bibinfo{note}{in Proceedings of the 2008 ACM/IEEE
  Conference on Supercomputing, Austin, Texas, 2008 (IEEE, New York, 2008)}.

\bibitem[{\citenamefont{Vary et~al.}(2009)\citenamefont{Vary, Maris, Ng, Yang,
  and Sosonkina}}]{Vary:2009qp}
\bibinfo{author}{\bibfnamefont{J.~P.} \bibnamefont{Vary}},
  \bibinfo{author}{\bibfnamefont{P.}~\bibnamefont{Maris}},
  \bibinfo{author}{\bibfnamefont{E.}~\bibnamefont{Ng}},
  \bibinfo{author}{\bibfnamefont{C.}~\bibnamefont{Yang}}, \bibnamefont{and}
  \bibinfo{author}{\bibfnamefont{M.}~\bibnamefont{Sosonkina}},
  \bibinfo{journal}{J. Phys. Conf. Ser.} \textbf{\bibinfo{volume}{180}},
  \bibinfo{pages}{012083} (\bibinfo{year}{2009}), \eprint{0907.0209}.

\bibitem[{\citenamefont{Maris et~al.}(2010{\natexlab{c}})\citenamefont{Maris,
  Sosonkina, Vary, Ng, and Yang}}]{Maris:2010aa}
\bibinfo{author}{\bibfnamefont{P.}~\bibnamefont{Maris}},
  \bibinfo{author}{\bibfnamefont{M.}~\bibnamefont{Sosonkina}},
  \bibinfo{author}{\bibfnamefont{J.~P.} \bibnamefont{Vary}},
  \bibinfo{author}{\bibfnamefont{E.~G.} \bibnamefont{Ng}}, \bibnamefont{and}
  \bibinfo{author}{\bibfnamefont{C.}~\bibnamefont{Yang}},
  \bibinfo{journal}{Procedia Comp. Sci.} \textbf{\bibinfo{volume}{1}},
  \bibinfo{pages}{97} (\bibinfo{year}{2010}{\natexlab{c}}).

\bibitem[{\citenamefont{Kortelainen and Lesinski}(2010)}]{Kortelainen:2010wb}
\bibinfo{author}{\bibfnamefont{M.}~\bibnamefont{Kortelainen}} \bibnamefont{and}
  \bibinfo{author}{\bibfnamefont{T.}~\bibnamefont{Lesinski}},
  \bibinfo{journal}{J. Phys.} \textbf{\bibinfo{volume}{G37}},
  \bibinfo{pages}{064039} (\bibinfo{year}{2010}), \eprint{1002.1321}.

\bibitem[{\citenamefont{Drut and Platter}(2011)}]{Drut:2011te}
\bibinfo{author}{\bibfnamefont{J.~E.} \bibnamefont{Drut}} \bibnamefont{and}
  \bibinfo{author}{\bibfnamefont{L.}~\bibnamefont{Platter}}
  (\bibinfo{year}{2011}), \eprint{1104.4357}.

\bibitem[{\citenamefont{Reinhard et~al.}(2006)\citenamefont{Reinhard, Bender,
  Nazarewicz, and Vertse}}]{Reinhard:2005nj}
\bibinfo{author}{\bibfnamefont{P.-G.} \bibnamefont{Reinhard}},
  \bibinfo{author}{\bibfnamefont{M.}~\bibnamefont{Bender}},
  \bibinfo{author}{\bibfnamefont{W.}~\bibnamefont{Nazarewicz}},
  \bibnamefont{and} \bibinfo{author}{\bibfnamefont{T.}~\bibnamefont{Vertse}},
  \bibinfo{journal}{Phys.Rev.} \textbf{\bibinfo{volume}{C73}},
  \bibinfo{pages}{014309} (\bibinfo{year}{2006}), \eprint{nucl-th/0510039}.

\end{thebibliography}

\end{document}